\documentclass{wileyarticle}

\articletype{Original Article}%
\artid{6384759}%
\jid{}%
\jvol{201}
\jyear{2017}%
\jissue{2}%
\jmonth{March}%
\doi{10.1155/2015/6384759}%
\copyeditor{}%
\startpage{1}%
\received{9 August 2017}
\revised{12 August 2017}
\accepted{19 August 2017}

\usepackage{lipsum}% this package is included to get dummy paragraphs for sample purpose.

\usepackage[T1]{fontenc}
\usepackage[utf8]{inputenc}
\usepackage{amsmath}
\usepackage{amsfonts}
\usepackage{amstext}
\usepackage{aas_macros}
\usepackage{graphicx}
\usepackage{natbib}
\renewcommand{\log}{\text{log}}

\usepackage[left=2cm,right=2cm,top=2cm,bottom=2cm]{geometry}

\begin{document}

\title{Radiative GRMHD simulations of accretion disks}

\author[]{W\l odek Klu\' zniak\footnotemark{2}}

\author[]{David Abarca\footnotemark{1}}

\author[]{Bhupendra Mishra}

\authormark{Klu\' zniak, Abarca, and Mishra}

%\footnotetext[1]{This is an example for first author footnote.}
%\footnotetext[2]{This is an example for second author footnote.}

\address[]{\orgname{Nicolaus Copernicus Astronomical Center, Bartycka 18}, \orgaddress{\city{ 00-716 Warszawa}, \country{Poland}}}

%%%%%\address[2]{\orgdiv{Original Division}, \orgname{Original University}, \orgaddress{\state{State name}, \country{Country name}}}

\corres{W{\l}odek Klu\'zniak, Centrum Astronomiczne im. Mikołaja Kopernika.
 % Copernicus Astronomical Center.
  \email{wlodek@camk.edu.pl}\\ 
  %Another corresponding author name, Brief address. \email{authorone@gmail.com}
}

%\presentaddress{Present address. Present address. Present address}

%%%%%%%%%%%%%%%%%%%%%%%%%%%%%%%%%%%%%%%%%%%%%%%%%%%%%%%%%%%%%%%%%%%%%%%%%%%%%%%
\begin{abstract}%
 We describe some recent results on the evolution of accretion disks around black holes and neutron stars obtained in magnetohydrodynamic (MHD) simulations in general relativity (GR) with the inclusion of radiation (GRRMHD).
\end{abstract}
%%%%%%%%%%%%%%%%%%%%%%%%%%%%%%%%%%%%%%%%%%%%%%%%%%%%%%%%%%%%%%%%%%%%%%%%%%%%%%

\jnlcitation{\cname{%
\author{W. Klu\'zniak},
\author{D. Abarca},  and 
\author{B. Mishra}} (\cyear{2017}), 
  \ctitle{Radiative GRMHD simulations of accretion disks},
  \cjournal{},%Q.J.R. Meteorol. Soc.,
  \cvol{}.}
  %2017;00:1--6}.

\maketitle
%%%%%%%%%%%%%%%%%%%%%%%%%%%%%%%%%%%%%%%%%%%%%%%%%%%%%%%%%%%%%%%%%%%%%%%%%%%%
%----------------------------------------------------------------------------
\section{Introduction}
%----------------------------------------------------------------------------
It is well known that matter
accreting onto a compact object from a binary companion has some angular momentum and forms an accretion disk as it falls towards the compact object. 
A mathematical solution of a geometrically thin alpha disk structure has been found by \citet{Shakura73}, here the word alpha stands for a phenomenological prescription of viscous stresses proportional to the pressure in the disk. For a detailed solution of the vertical structure of the alpha disk, including the velocity field, see \citet{2000astro.ph..6266K}. It is now thought that the dissipative stresses in hot accretion disks are a result of the magneto-rotational instability (MRI), which leads to magnetic field amplification on the orbital timescale \citep{Balbus92}. MRI only operates in those parts of the flow in which angular frequency decreases outwards, a condition certainly satisfied in GR for thin disks around black holes (i.e., in the Kerr metric).

Recent advances allowed radiation to be included in GRMHD codes, and we report some results obtained in our group with two such codes, both treating radiation in the M1 closure scheme: the \textsc{Koral} code \citep{sadowski+m1}, and the \textsc{Cosmos++} code \citep{Anninos05,Fragile14}.

%------------------------------------------------------------------------- 
  \begin{figure*}
 	\centering
 	\SPIFIG
  {\begin{tabular}{cc}
		\includegraphics[width=2\columnwidth]{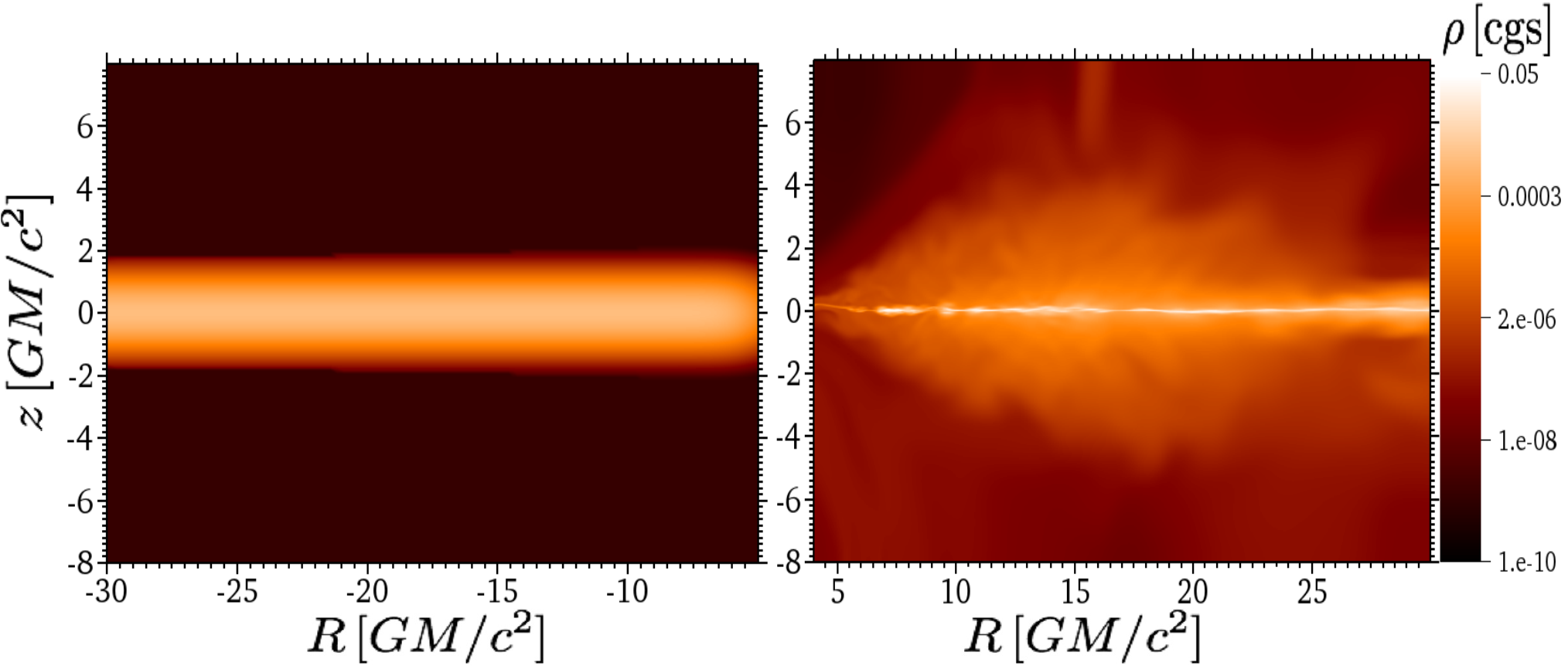}           
 	\end{tabular}}
{ \caption{Mass-density plot of three dimensional radiation pressure dominated simulation illustrating its thermal instability. The left panel shows initial setup and the right panel shows collapsed final stage of the disk. Figure reproduced from \citet{Mishra16}.
 \label{fig:collapse}}}
 \end{figure*}
%------------------------------------------------------------------------- 

%----------------------------------------------------------------------------
\section{Governing Equations}
%----------------------------------------------------------------------------
The equations of GRRMHD, can be written in their conservative
form as
\begin{align}
\nabla_\mu (\rho u^\mu) &= 0, \\
\nabla_\mu T^\mu\phantom{}_\nu &= G_\nu, \\
\nabla _\mu R^\mu\phantom{}_\nu &= -G_\nu, \\
\nabla_\mu(nu^\mu) &= \dot{n},
\end{align}

Here, $\rho$ is the gas density in the comoving fluid frame, $u^\mu$ is the 
gas four-velocity, $T^\mu\phantom{}_\nu$ is the MHD stress-energy tensor 
The radiation stress-energy tensor, 
$R^\mu\phantom{}_\nu$, completed using the $M_1$ closure scheme
which assumes there is a frame in which the 
radiation is isotropic (that frame being usually quite distinct from the comoving fluid frame), is coupled to the gas stress-energy tensor by
the radiation four-force, $G_\nu$. Electron scattering and bremsstrahlung
opacities as well as photon conserving Comptonization are included in the codes \citep{sadowski+comp}.
%------------------------------------------------------------------------- 
 \begin{figure*}
	\centering
 	\SPIFIG
             {\begin{tabular}{cc}
 		\includegraphics[scale=.15]{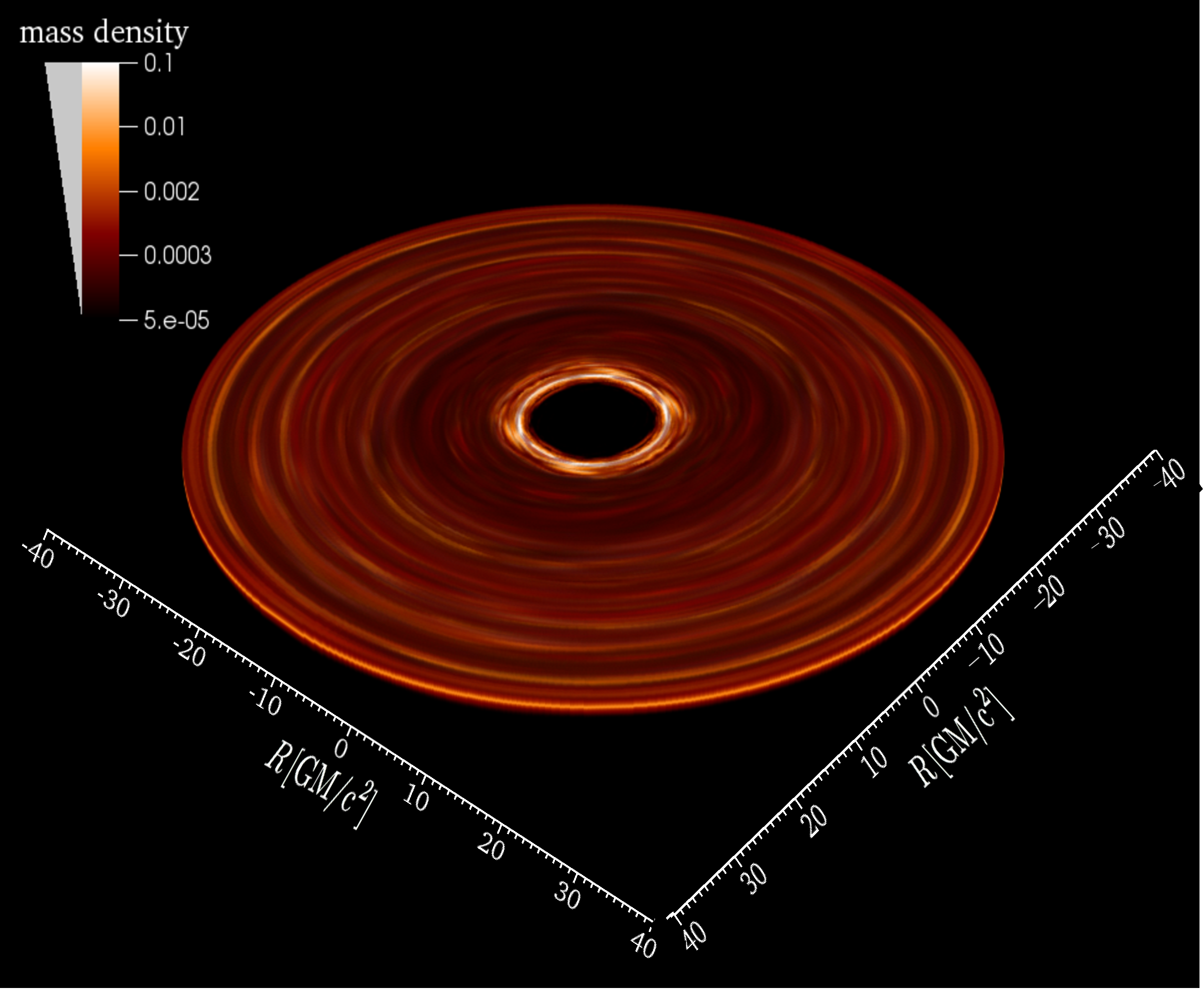} 
      	\end{tabular}}           
{ \caption{Volume plot of mass density for a radiation pressure dominated thin disk simulation illustrating the Lightman-Eardley instability.  Figure reproduced from \citet{Mishra16}.
 \label{fig:leardley}}}
 \end{figure*}
%------------------------------------------------------------------------- 

%----------------------------------------------------------------------------
\section{Stability of geometrically thin acretion disks}
%----------------------------------------------------------------------------
For mass accretion rates yielding a luminosity, $L$, exceeding $\sim 0.1$ of the Eddington luminosity $L_\mathrm{Edd}$, the inner parts of the disk have been found to be dominated by radiation pressure \citep{Shakura73}. As found by \citet{Shakura76} in a linear perturbation analysis, such disks are unstable to a thermal runaway \citep[see also][]{2012A&A...538A.148C,2011MNRAS.415.2319L}.

With the \textsc{Cosmos++} code, we have performed global GRRMHD 3D simulations of thin accretion disks \citep[][the reader is referred to this reference for all details]{Mishra16}, and in agreement with previous shearing-box simulations we have found the radiation-pressure dominated disks to collapse on the thermal timescale. Figure~\ref{fig:collapse} shows the initial and final disk configuration.

Figure~\ref{fig:leardley} shows a perspective view of the end state of the same simulation. Note the ring-like structures, especially the one close to the inner edge of the accretion disk (which is close to the ISCO, the innermost stable circular orbit of test particles). We interpret this as resulting from the viscous instability, discussed in the remarkable paper by \citet{Lightman74}, who
showed that viscous, radiation-pressure dominated disks suffer an instability, formally corresponding to a negative coefficient in the diffusion equation, leading to the formation of regions of enhanced density. As predicted by \citet{Lightman74}, we find the dissipative stresses to be anticorrelated with density. This is the first time in more than four decades that the effect has been reported in numerical simulations, as nobody had ever run simulations of thin accretion disks with the inclusion of radiation prior to the \citet{Mishra16} work.
A gas-pressure dominated disk at lower luminosities (not shown) has been found to be stable throughout the duration of the simulation, confirming that the reported instabilities are not of numerical origin. 
%------------------------------------------------------------------------- 
 \begin{figure*}
 	\centering
 	\SPIFIG
{ 	\begin{tabular}{cc}
 		$\log_{10} \hat{E}\, \left[\text{erg/cm}^3\right]$\hfill $
 		\log_{10}\rho\, \left[\text{g/cm}^3\right]$& 
 		$\log_{10} \hat{E}\, \left[\text{erg/cm}^3\right]$\hfill $
 		\log_{10}\rho\, \left[\text{g/cm}^3\right]$ \\
 		\includegraphics[scale=.25]{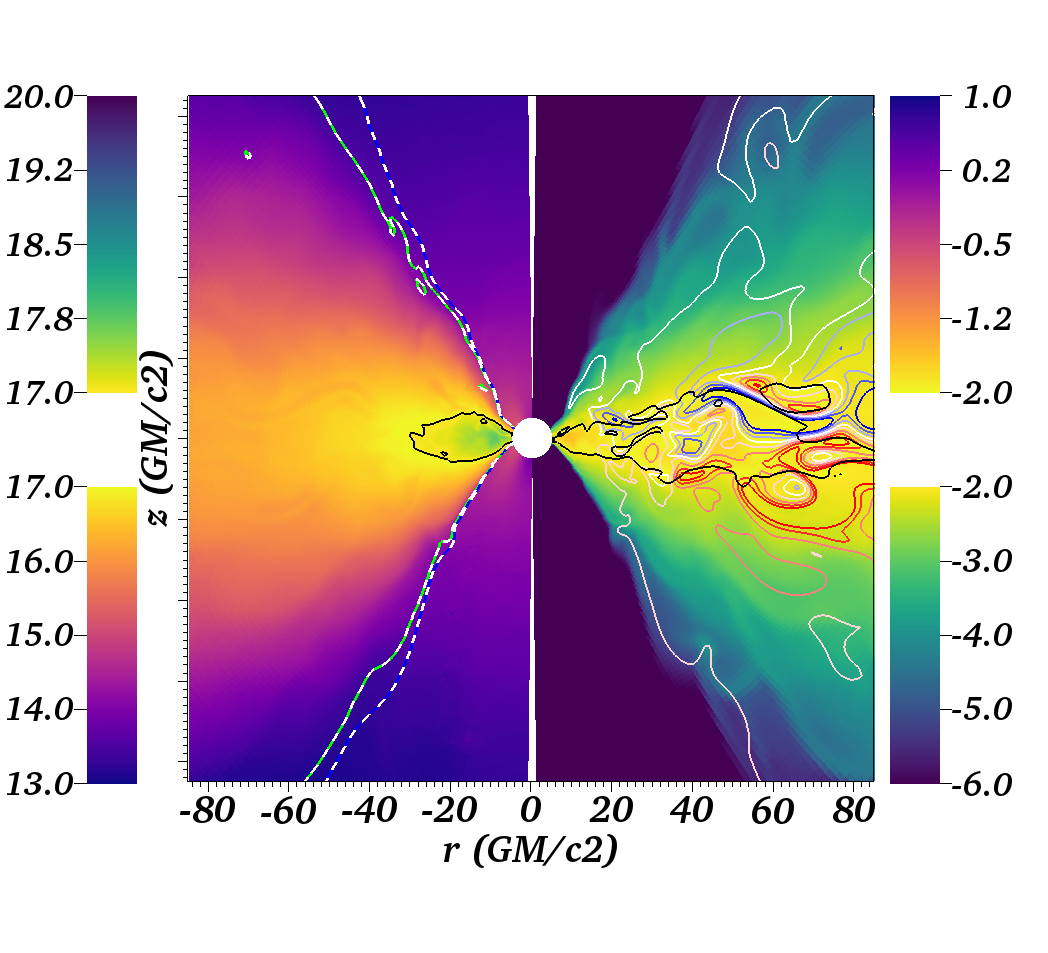} &
 		\includegraphics[scale=.25]{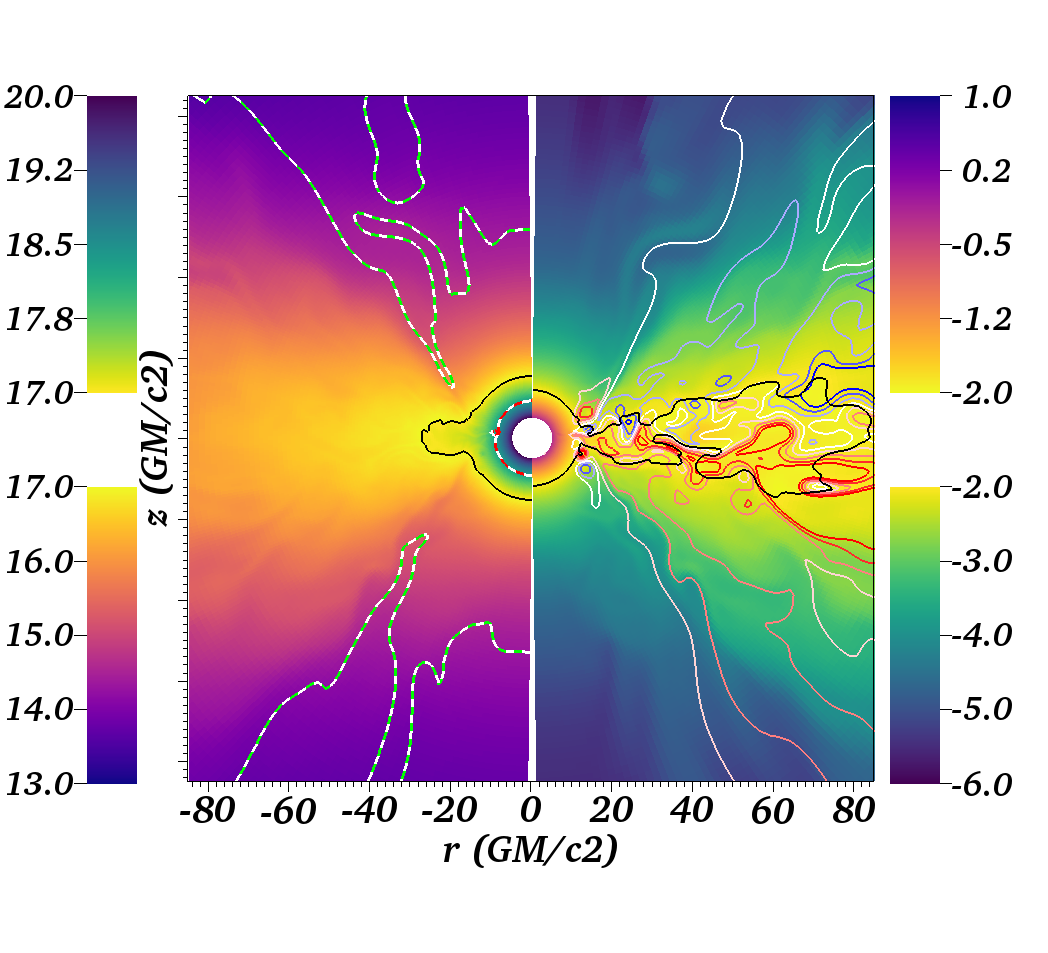}
 	\end{tabular}}
{ 	\caption{Here we show snapshots of the 2.5D axisymmetric
    simulations of super-Eddington accretion onto a black hole (left
    panel) and a neutron star (right panel). The left side of each
    panel shows the log of radiative energy density in the fluid
    frame, and the right side shows the log of rest mass density.  We
    use two color scales. A black contour marks the division between
    the regions of different color scales. Dashed contours on the left
    panel show the relativistic Bernoulli number. Red is $Be = -0.05$,
    green is $Be=0$, and blue is $Be=0.05$.  Solid contours on the
    right show the poloidal magnetic field. Blue and  red and used to
    distinguish left and right handed loops. 
 \label{fig:bhframes}}}
 \end{figure*}
%------------------------------------------------------------------------- 

%------------------------------------------------------------------------- 
 
 \begin{figure*}
 	\centering
 	\SPIFIG
 {\begin{tabular}{cc}
     \includegraphics[scale=.5]{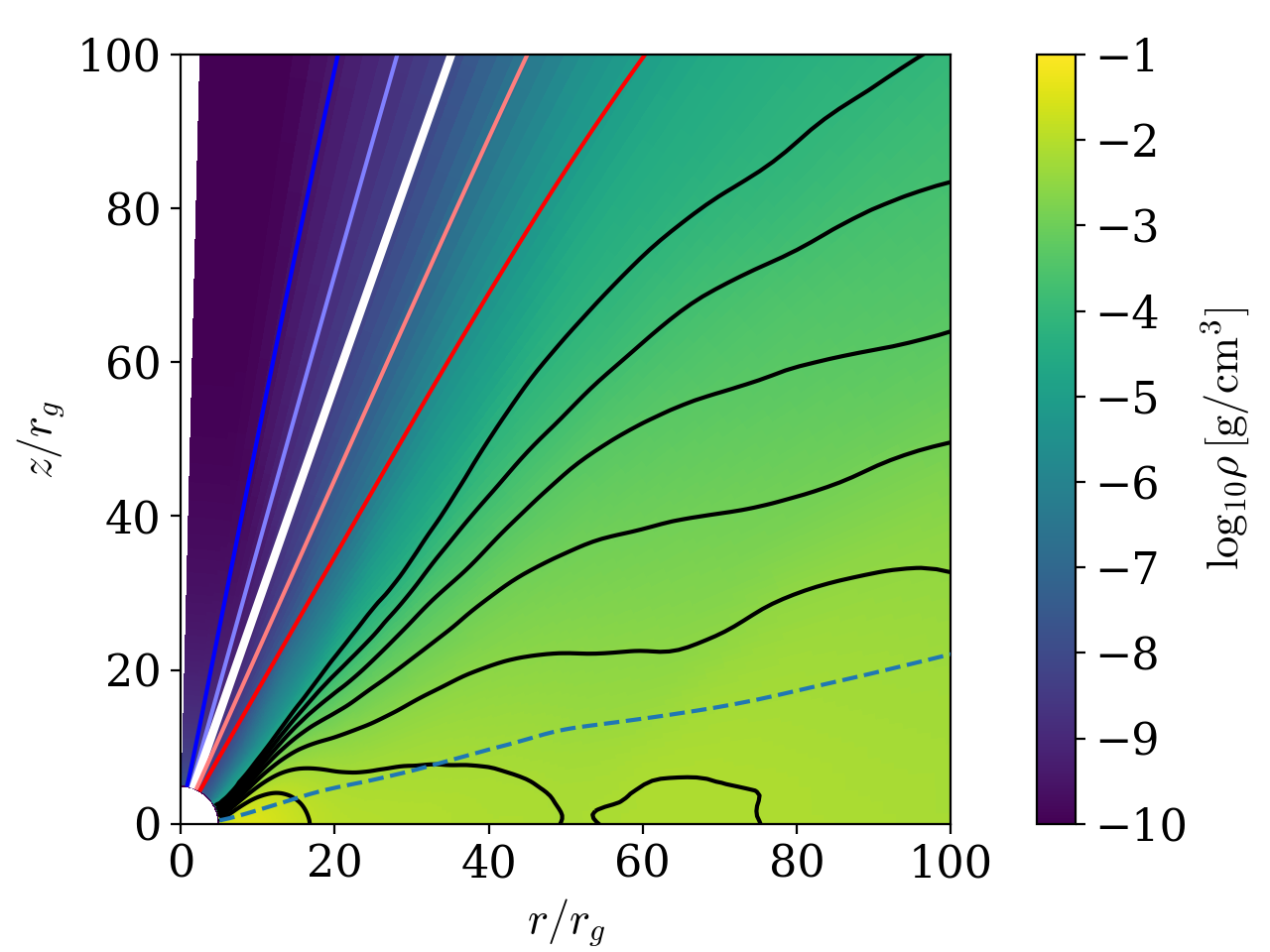} & \includegraphics[scale=.5]{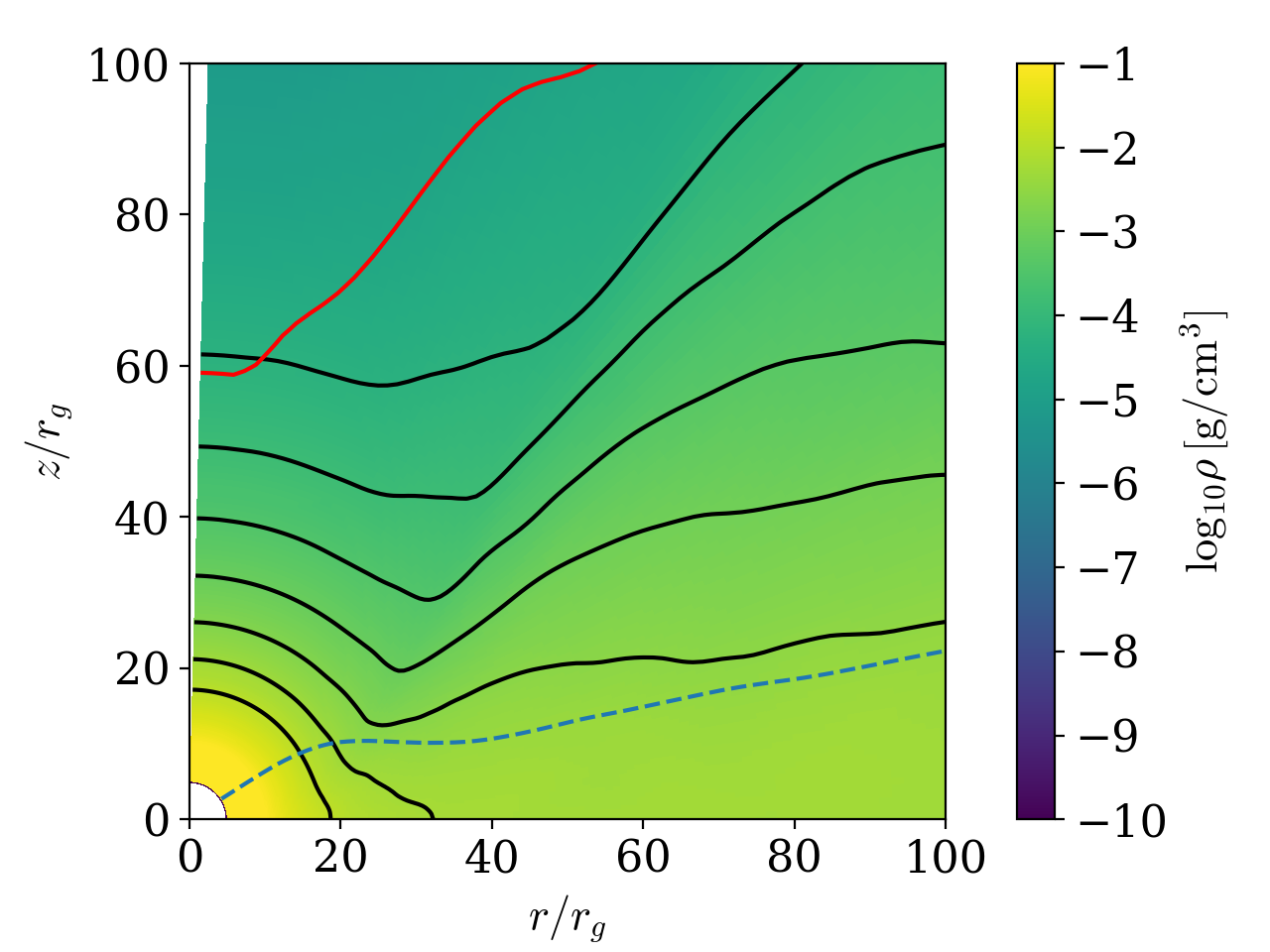} 
%       figs/bh_10mdot} & \includegraphics[scale=.5]{figs/ns_10mdot} 
 	\end{tabular}}
{ \caption{Here we show the time averaged structure of the black hole
    simulation (left) and the neutron star simulation (right).  The
    colormap shows density. Log-spaced pressure contours are shown  in
    black. The colored contours correspond to optical depth. From blue
    to red the values correspond to $\tau = 0.01, 0.1, 1, 10, 100$.
    The dashed blue line shows the scale height of the disk.
 \label{fig:avg}}}
 \end{figure*}
 %-------------------------------------------------------------------------

 %----------------------------------------------------------------------------
\section{Ultraluminous X-ray sources (ULX)}
%----------------------------------------------------------------------------
One particular class of accreting objects 
which has gained interest in recent years are ultraluminous
X-ray sources (ULXs). These are X-ray bright objects with observed luminosities 
up to  $~10^{41}$ ergs s$^{-1}$. 
Currently, the leading explanation for ULXs is beamed emission from 
accretion in an X-ray binary
\citep{king+01}, implying that near- or super-Eddington accretion is responsible for the large observed luminosities. 
In particular, a  set of three such  
objects were observed which reveal X-ray pulsations 
with a period on the order of one second \citep{bachetti+14,furst+16,israel+17a,israel+17b}
indicating the presence of pulsars as the accreting objects.
It can now be said with some certainty that a large fraction of 
ULXs are accreting neutron stars \citep{king+17}.

We present preliminary results of simulations (Abarca et al., in preparation) of super-Eddington accretion onto a netron star, which may be relevant to ULXs, using the GRRMHD code 
\textsc{Koral} in a 2.5D implementation \citep{sadowski+dynamo} allowing axisymetric accretion disk simulations 
to be run for long durations without depleting the magnetic field due to 
turbulent dissipation,which normally occurs in axisymmetric
simulations of MRI. The effects of a stellar magnetic field have been ignored, and the neutron starr calculations are carried out in the Schwarzschild metric.
Gas accreting onto the outer layers of a neutron star is expected to slow down and release its kinetic energy \citep{ss+86},
which can be
converted into radiation, as in the study of \citep{kluzniak+91}, or transferred to outflowing gas.

\subsection{Initial conditions and boundary conditions}
We initialize our accretion disk in a standard way by starting with an equilibrium
torus near the black hole as given in \citep{penna+13}. The torus is then threaded
with a weak magnetic field of alternating polarity. The total pressure is then distributed
between gas and radiation assuming local thermal equilibrium. Once the simulation
starts, the MRI quickly develops turbulence and accretion begins.

For the neutron star-like case, we implement a reflective boundary. 
The reflective boundary is set
up so that the reconstructed radial velocity at the inner boundary is zero. We also set
the perpendicular velocities $u^\theta, u^\phi$ to zero. The inner boundary is then set to 
only exchange momentum so that we can be sure no mass or energy leaves or enters
the domain. For comparison purposes, we also run a black hole-like simulation in the Schwarzschild metric with inflow boundary conditions.
 
 \subsection{Preliminary Results}
 
 We have run two axisymmetric simulations, one with a reflective inner boundary,
 and one with a black hole-like inner boundary, both at radius $r=5GM/c^2$. 
 Snapshots of both are shown in Figs.~\ref{fig:bhframes}.
 
 In the neutron star simulation gas cannot 
 pass through the inner boundary, so it accumulates into a hot, dense atmosphere.
 In addition gas is blown off the outer edges of the atmosphere forming a
 dense outflow.  The radiation energy density is very high, but it is not clear how much of this 
 radiation escapes to infinity. The gas with positive Bernoulli number, of which there is a significant amount of, can still absorb or emit the radiation in the funnel region before
 it reaches infinity.

 In Fig.~\ref{fig:avg} we show the time averaged spatial structure of both disks. The
 black hole-like simulation is as expected. Super-Eddington accretion leads to a thick disk with 
 strong outflows and an optically thin funnel region reaching all the way down to the 
 inner boundary. The accretion flow is somewhat sub-Keplerian, which manifests as a 
 weak radial pressure gradient. We see a different picture for the reflective boundary. 
 First, as is also seen in Fig.~\ref{fig:bhframes} a large amount of gas is deposited 
 in an atmosphere around the inner boundary. In addition, a large amount of gas is
 ejected and the entire domain is filled with an optically thick outflow. The photosphere 
 extends practically to the edge of the simulation domain at 100 stellar radii. The last visible
 contour of optical depth shows an optical depth of about 100 at radius sixty. It is hard to make a statement about the radiative properties as viewed from infinity. However, it is certain that the neutron star surface would be unobservable in this case.
 A measurement of the radiative flux at the edge of the simulation domain shows a flux which
 is locally super-Eddington when viewed along the poles. 
 
 Research supported in part by the Polish NCN grant 2013/08/A/ST9/00795

\section*{References}

\nocite{*}
\bibliography{pWKvaradero}%

%%\begin{biography}{\includegraphics[width=5pc,height=5pc]{blankfig.eps}}{\textbf{Author Name.} This is author biography text. This is author biography text. This is author biography text. This is author biography text. This is author biography text. This is author biography text. This is author biography text. This is author biography text. This is author biography text. This is author biography text.}
%\end{biography}

\end{document}